\documentclass[aps,prl,twocolumn,showpacs,preprintnumbers,amsmath,amssymb]{revtex4-1}

 \def\ep{{\epsilon}}

 \def\frac#1#2{{#1\over #2}}

 \def\s{\sqrt}

\def\be{\begin{equation}}
\def\ee{\end{equation}}
\def\ba{\begin{eqnarray}}
\def\ea{\end{eqnarray}}

 \def\de{\partial}

 \def\f {\frac}
 \def\ti{\tilde}

 \def\ddd{\cdot\cdot\cdot}
 \def\no{\nonumber \\}

 \def\la{\langle}
 \def\lb{\rangle}
 \def\ep{\epsilon}

\usepackage{color}
\usepackage{graphicx}
\usepackage{dcolumn}
\usepackage{bm}

\begin{document}

\title{Holographic Entanglement of Purification from
Conformal Field Theories}
YITP-18-125 ; IPMU18-0203
\author{Pawel Caputa$^{a}$, Masamichi Miyaji$^{a}$,   \\
Tadashi Takayanagi$^{a,b}$ and Koji Umemoto$^{a}$}

\affiliation{$^a$Center for Gravitational Physics,\\
Yukawa Institute for Theoretical Physics,
Kyoto University, \\
Kitashirakawa Oiwakecho, Sakyo-ku, Kyoto 606-8502, Japan}

\affiliation{$^{b}$Kavli Institute for the Physics and Mathematics
 of the Universe (WPI),\\
University of Tokyo, Kashiwa, Chiba 277-8582, Japan}

\date{\today}

\begin{abstract}
We explore a conformal field theoretic interpretation of the holographic entanglement of purification,
which is defined as the minimal area of entanglement wedge cross section. We argue
that in AdS$_3/$CFT$_2$, the holographic entanglement of purification agrees with
the entanglement entropy for a purified state, obtained from a special Weyl transformation, called path-integral optimizations.
By definition, this special purified state has the minimal path-integral complexity. We confirm this claim in several examples.
\end{abstract}

\maketitle


{\bf 1. Introduction}

Quantum entanglement provides us a key to understand how
gravity emerges from field theories.
This is manifest in the Anti de-Sitter space(AdS)/conformal field theory(CFT)
correspondence (or holography) \cite{Ma}.
In AdS/CFT, the entanglement entropy, the unique measure of quantum entanglement for pure states,
is computed as a minimal area in AdS \cite{RT,HRT}.

For mixed states, entanglement entropy can measure neither quantum entanglement nor correlations between two subsystems. Recently, it was conjectured that a quantity called entanglement of purification
\cite{EP}, which is a good measure of total correlations, has a simple geometric interpretation in AdS/CFT, called the holographic entanglement of purification \cite{UT,Nguyen:2017yqw}. This quantity is a natural extension of entanglement entropy for mixed states. An extension to multi-partite correlations is also given in \cite{Umemoto:2018jpc}.

The entanglement of purification (EoP), written as $E_P(\rho_{AB})$, is a measure of correlations between two subsystems $A$ and $B$ for a mixed state $\rho_{AB}$ and is defined as follows \cite{EP}. Consider a purification of $\rho_{AB}$,
given by $|\Psi\lb_{A\ti{A}B\ti{B}}$ , by enlarging the Hilbert space as
${\cal H}_A\otimes {\cal H}_B \to {\cal H}_A\otimes {\cal H}_B \otimes {\cal H}_{\ti{A}}
\otimes {\cal H}_{\ti{B}}$ such that
\be
\rho_{AB}=\mbox{Tr}_{\ti{A}\ti{B}}\left[|\Psi\lb_{A\ti{A}B\ti{B}}\la\Psi|_{A\ti{A}B\ti{B}}\right].
\label{condw}
\ee
Among infinitely many different choices of the purifications $|\Psi\lb_{A\ti{A}B\ti{B}}$,
the EoP is defined by minimizing the entanglement entropy
$S_{A\ti{A}}=-\mbox{Tr}[\rho_{A\ti{A}}\log\rho_{A\ti{A}}]$ as
\ba
E_P(\rho_{AB})=\min_{|\Psi\lb_{A\ti{A}B\ti{B}}}[S_{A\ti{A}}].  \label{EoPdef}
\ea

The holographic entanglement of purification (HEoP) $E_W(\rho_{AB})$ is defined by the area of the minimal cross section of entanglement wedge, denoted by $\Sigma^{min}_{AB}$:
\be
E_W(\rho_{AB})=\frac{A(\Sigma^{min}_{AB})}{4G_N}, \label{heop}
\ee
and the equality $E_W(\rho_{AB})=E_P(\rho_{AB})$ was conjectured in \cite{UT,Nguyen:2017yqw}
based on quantum information theoretic properties.
In this paper, we focus on static backgrounds and always take a canonical time slice in the AdS.

The entanglement wedge is the region inside the AdS which is dual to the density matrix $\rho_{AB}$
\cite{EW1,EW2,EW3}. Therefore, the HEoP (\ref{heop}) is quite fundamental to understand how the
geometry in gravity corresponds to density matrices in CFTs. It also has an important meaning
in the bit threads interpretation of AdS/CFT \cite{Freedman:2016zud,Hubeny:2018bri,Cui:2018dyq,Agon:2018lwq}.

Nevertheless, direct comparisons between the HEoP and EoP have not been done so far
mainly because the minimization procedure in (\ref{EoPdef}) is very hard in quantum field theories.
Refer to \cite{Nguyen:2017yqw,Zal,Bhattacharyya:2018sbw} for numerical lattice calculations of EoP in free field theories.
Also in \cite{Hirai:2018jwy,Kudler-Flam:2018qjo,Tamaoka:2018ned}, connections between $E_W(\rho_{AB})$
and quantities other than the EoP have been proposed. For other recent progresses on the HEoP, refer to \cite{HEP1,HEP2,HEP3,HEP4,HEP5,HEP6,HEP7,HEP8,HEP9,HEP10,HEP11,HEP12,HEP13}

In this letter, we would like to present a direct comparison between the HEoP $E_W(\rho_{AB})$ and the entanglement entropy $S_{A\ti{A}}$ for a special class of purification $|\Psi\lb_{A\ti{A}B\ti{B}}$, obtained by the path-integral optimization \cite{Caputa:2017urj,Bhattacharyya:2018wym}. We will focus on several examples in AdS$_3/$CFT$_2$ and will find both quantities always agree with each other in the regime
of validity of our computations.\\

{\bf 2. Path-Integral Optimization}

Consider a two dimensional (2D) CFT on a Euclidean flat space $R^2$, which is described by the complex coordinate $(w,\bar{w})=(\xi+i\tau,\xi-i\tau)$ with the metric
\be
ds^2=dwd\bar{w}=d\tau^2+d\xi^2.  \label{flat}
\ee
The theory is invariant under the Weyl transformation of the metric:
\be
ds^2=e^{2\phi(\tau,\xi)}(d\tau^2+d\xi^2),  \label{Weyl}
\ee
where $\phi(\tau,\xi)$ is any function.

The path-integral optimization \cite{Caputa:2017urj} is a special choice of Weyl transformation
which (i) preserves the quantum state $|\Psi\lb$ at a particular time $\tau=\tau_0$ and (ii) minimizes the path-integral complexity $C_L[\phi]$ (defined below).

Since the wave functional of $|\Psi\lb$ can be computed by the Euclidean path-integral on the lower half plane $-\infty<\tau\leq \tau_0$, the first condition (i) is given by the boundary condition
\be
e^{2\phi(\tau_0,\xi)}=1. \label{conda}
\ee
The path-integral complexity is defined by the Liouville action
\be
C_L[\phi]=\frac{c}{24\pi}\int^{\tau_0}_{-\infty} d\tau\int^\infty_{-\infty} d\xi \left[(\de_{\tau}\phi)^2+(\de_{\xi}\phi)^2+\frac{e^{2\phi}}{\ep^2} \right], \label{pco}
\ee
where $\ep$ is the UV cut off or equally lattice spacing and $c$ is the central charge.
The reason why we identify $C_L[\phi]$
with the complexity in path-integrations is because the path-integral measure is proportional to
$e^{C_L[\phi]}$ due to the conformal anomaly \cite{Po}. This quantity $C_L[\phi]$ provides a field theoretic counterpart of holographic complexity \cite{Susskind} as explained in \cite{TaC}.

In this argument, we consider the discretization of path-integral such that each cell has the
area $\ep^2$. The original flat metric (\ref{flat}) corresponds to a square lattice with lattice spacing $\Delta\xi=\Delta \tau=\ep$. The optimization which changes the metric such that $e^{2\phi}\leq 1$
means coarse-graining lattice sites such that  $\Delta\xi=\Delta \tau=\ep\cdot e^{-\phi}$.

The minimization of $C_L[\phi]$ can be found by solving the Liouville equation
$(\de^2_\tau+\de^2_\xi)\phi=e^{2\phi}/\ep^2$.
Choosing $\tau_0=-\ep$ using the time translational symmetry, we find the solution
which satisfies the condition (\ref{conda})
\be
e^{2\phi}=\frac{\ep^2}{\tau^2}. \label{newc}
\ee
This describes a hyperbolic space H$_2$.
Note that in \cite{Caputa:2017urj} we had $e^{2\phi}=1/z^2$ because we defined $z=-\ep$ with the rescaling $e^\phi\to \ep\cdot e^{\phi}$. This is the simplest example of the path-integral optimization. Even though we
optimized a pure state $|\Psi\lb$, it is straightforward to extend the formulation such that we optimize a given mixed state density matrix.  \\

{\bf 3. Optimization of Single Interval}

\begin{figure}
  \centering
 \includegraphics[width=8cm]{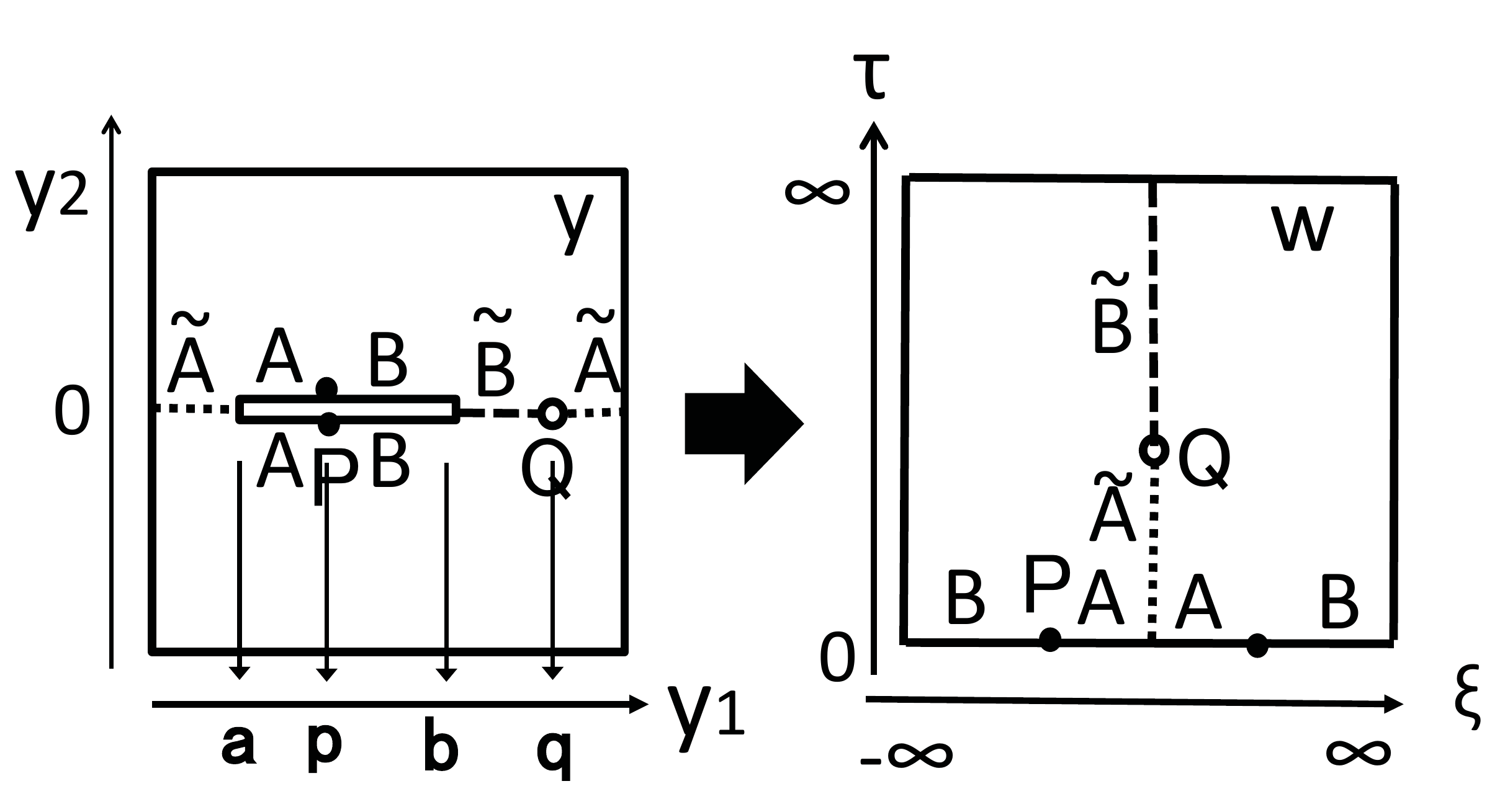}
 \caption{Conformal transformation between a complex plane with one slit and an upper half plane.}
\label{fig:ConformalP}
\end{figure}

Consider the case where the subsystem $AB(\equiv A\cup B)$ is given by a single interval in a vacuum.
We parameterize the subsystem $A$ and $B$ as follows:
\ba
A=[a,p],\ \ \ B=[p,b],\ \ \ (-\infty< a<p<b<\infty),
\ea
where they share the point $P$ given by $y=p$.
The reduced density matrix $\rho_{AB}$ is given by the Euclidean path-integral over a complex plane with a slit along $AB$, whose coordinate is denoted by $y$. Next we consider a purification $|\Psi\lb_{A\ti{A}B\ti{B}}$
of $\rho_{AB}$ by introducing the subsystem $\ti{A}$ and $\ti{B}$ as in Fig.\ref{fig:ConformalP} such that they share the point $Q$ parameterized by $y=q$.

The original definition of EoP is the minimum of $S_{A\ti{A}}$ against any purifications (\ref{EoPdef}). Here we would like to restrict to a class of
purification which is realized by the Weyl transformation (\ref{Weyl}) and focus on the one given by the path-integral optimization (i.e. the one which minimizes $C_L[\phi]$). The purification condition
(\ref{condw}) is equivalent to the requirement (\ref{conda}).

To find the optimization, let us map the $y$-space into an upper half-plane ($w$-plane) by
\ba
w=\s{\f{y-a}{b-y}}.
\ea
The point $P$ and $Q$ are mapped into $w_P=-\s{\frac{p-a}{b-p}}$ and
$w_Q=i\s{\frac{q-a}{q-b}}$.

Since the path-integral optimization on the upper half plane is given by the hyperbolic space (\ref{newc}), the final optimized metric reads
\be
ds^2=\frac{\ep^2}{\tau^2}\cdot dwd\bar{w}=
\frac{\ep^2}{\tau^2}\cdot\frac{(b-a)^2}{4|b-y|^3|y-a|}\cdot dyd\bar{y}\equiv e^{2\ti{\phi}}\cdot dyd\bar{y}.
\label{metsop}
\ee
The coordinate $\tau$ takes values in the range $-\infty<\tau<-\delta_1$, where $\delta_1(>0)$ is an appropriate
cut off so that we maintain the condition (\ref{condw}). However, the detail of this regularization is not important in the coming calculations and refer to the appendix A for more explanations.
In this setup, we have
$e^{2\ti{\phi}_P}=1$, which follows from the condition (\ref{flat}), and we find $e^{2\ti{\phi}_Q}=\frac{\ep^2(b-a)^2}{4(q-a)^2(q-b)^2}$, where $\ti{\phi}_{P}$
and  $\ti{\phi}_{Q}$ are the value of $\ti{\phi}$ at $P$ and $Q$.

The entanglement entropy $S_{A\ti{A}}=S_{B\ti{B}}$ is found to be
\ba
S_{A\ti{A}}&=&\frac{c}{3}\log\left(\frac{q-p}{\ep}\right)+\frac{c}{6}\ti{\phi}_{P}+\frac{c}{6}\ti{\phi}_{Q} \no
&=&\frac{c}{6}\log\left[\frac{(b-a)(q-p)^2}{2\ep(q-a)(q-b)}\right],
\ea
where in the first line we performed a scale transformation of the standard formula of \cite{HLW}.
Remember that the entanglement entropy can be found from the two point function of twist operators 
in the replica method and this transforms in a standard way under the Weyl transformation. Then we minimize $S_{A\ti{A}}$ with respect to $q$, leading to
\ba
&& S^{min}_{A\ti{A}}= \frac{c}{6}\log\left[\frac{2(p-a)(b-p)}{\ep(b-a)}\right], \label{eopmin}\\
&& \  \ \mbox{at}\ \ \  q=\frac{2ab-(a+b)p}{a+b-2p}.  \label{qqqw}
\ea

\begin{figure}
  \centering
 \includegraphics[width=6cm]{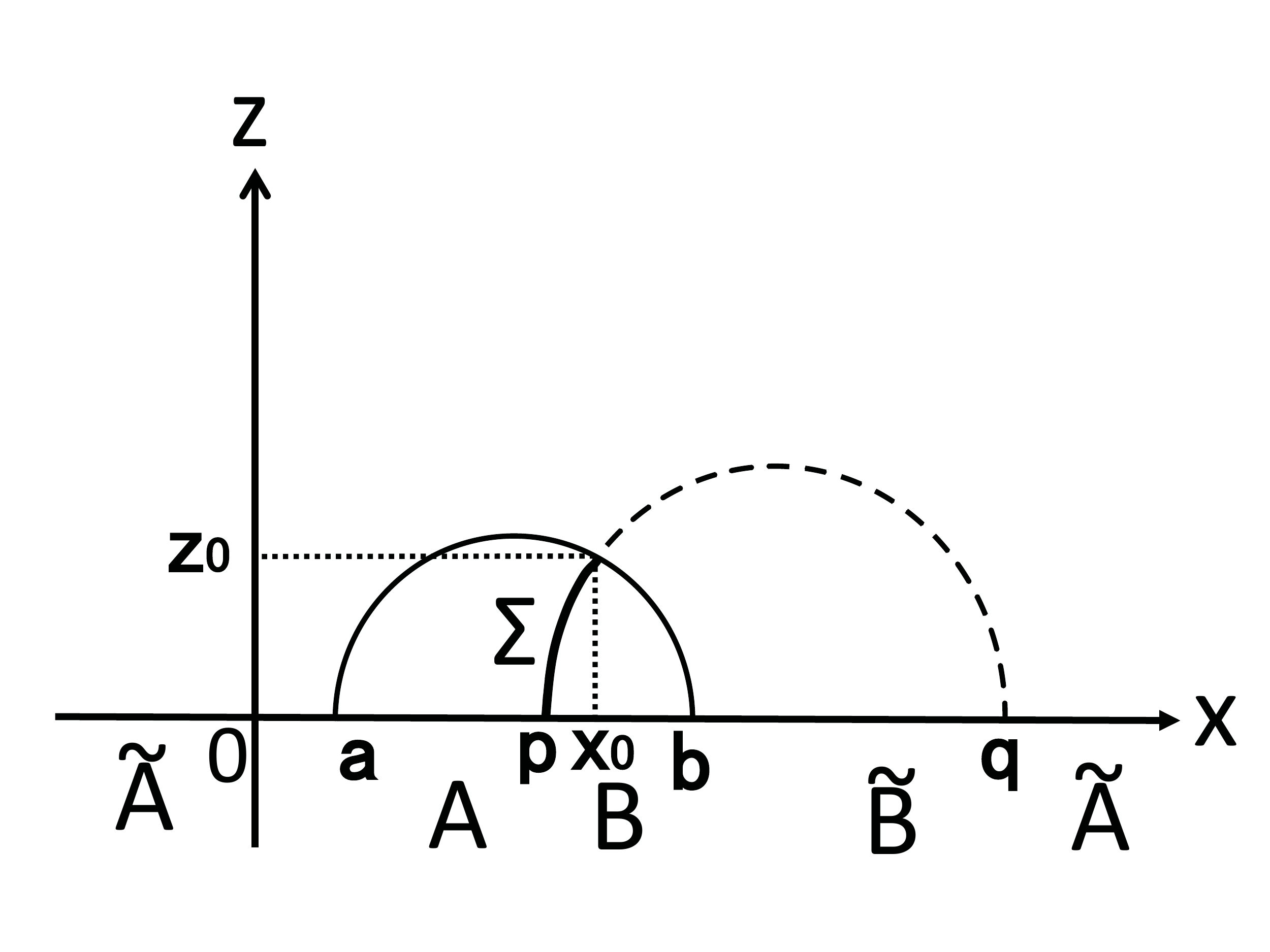}
 \caption{Calculation of Holographic EoP. The length of $\Sigma$ (thick curve in the above picture) multiplied by $\frac{1}{4G_N}=\frac{c}{6}$ gives the holographic EoP (\ref{heop}).}
\label{fig:ConformalPH}
\end{figure}

The holographic EoP is computed from the length of the entanglement wedge cross section $\Sigma^{min}_{AB}$ in the hyperbolic space $ds^2=(dx^2+dz^2)/z^2$, which is a time slice of Poincare AdS$_3$. The geodesic curve $\Sigma^{min}_{AB}$ is a part of a circle which connects $P:(x,z)=(p,\ep)$ and $Q:(x,z)=(q,\ep)$, as depicted in Fig.\ref{fig:ConformalPH}.
The minimal length condition requires that $\Sigma^{min}_{AB}$ is perpendicular to the circle $\left(x-\frac{a+b}{2}\right)^2+z^2=\frac{(b-a)^2}{4}$ at their intersection point $(x,z)=(x_0,z_0)$. This fixes the value of $q$ as (\ref{qqqw}) and we obtain $x_0=\frac{pq-ab}{p+q-(a+b)}$ and
$z_0=\s{\frac{(b-a)^2}{4}-\left(x_0-\frac{a+b}{2}\right)^2}$.
Its length $A(\Sigma^{min}_{AB})$ is computed as
\ba
\frac{q-p}{2}\int^{z_0}_{\ep}\frac{dz}{z\s{\frac{(q-p)^2}{4}-z^2}}
=\log\left[\frac{2(p-a)(p-b)}{\ep(b-a)}\right].\nonumber
\ea
Thus the holographic EoP (\ref{heop}) perfectly matches with (\ref{eopmin}).

It is straightforward to extend the above agreement between $E_W(\rho_{AB})$ and $S_{A\ti{A}}$
to the finite temperature case by performing a conformal transformation $y=e^{\frac{2\pi}{\beta}\zeta}$.
Similarly, we can extend it to the finite size case.\\

{\bf 4. Optimization of Double Intervals}

As the second class of examples, we consider the case where the subsystems $A$ and $B$ are
mutually disconnected intervals on an infinite line. Though we focus on the vacuum state
in a two dimensional CFT below, it is straightforward to extend to the finite temperature setup
by the conformal map $y=e^{\frac{2\pi}{\beta}\zeta}$ as before.

\begin{figure}
  \centering
 \includegraphics[width=8cm]{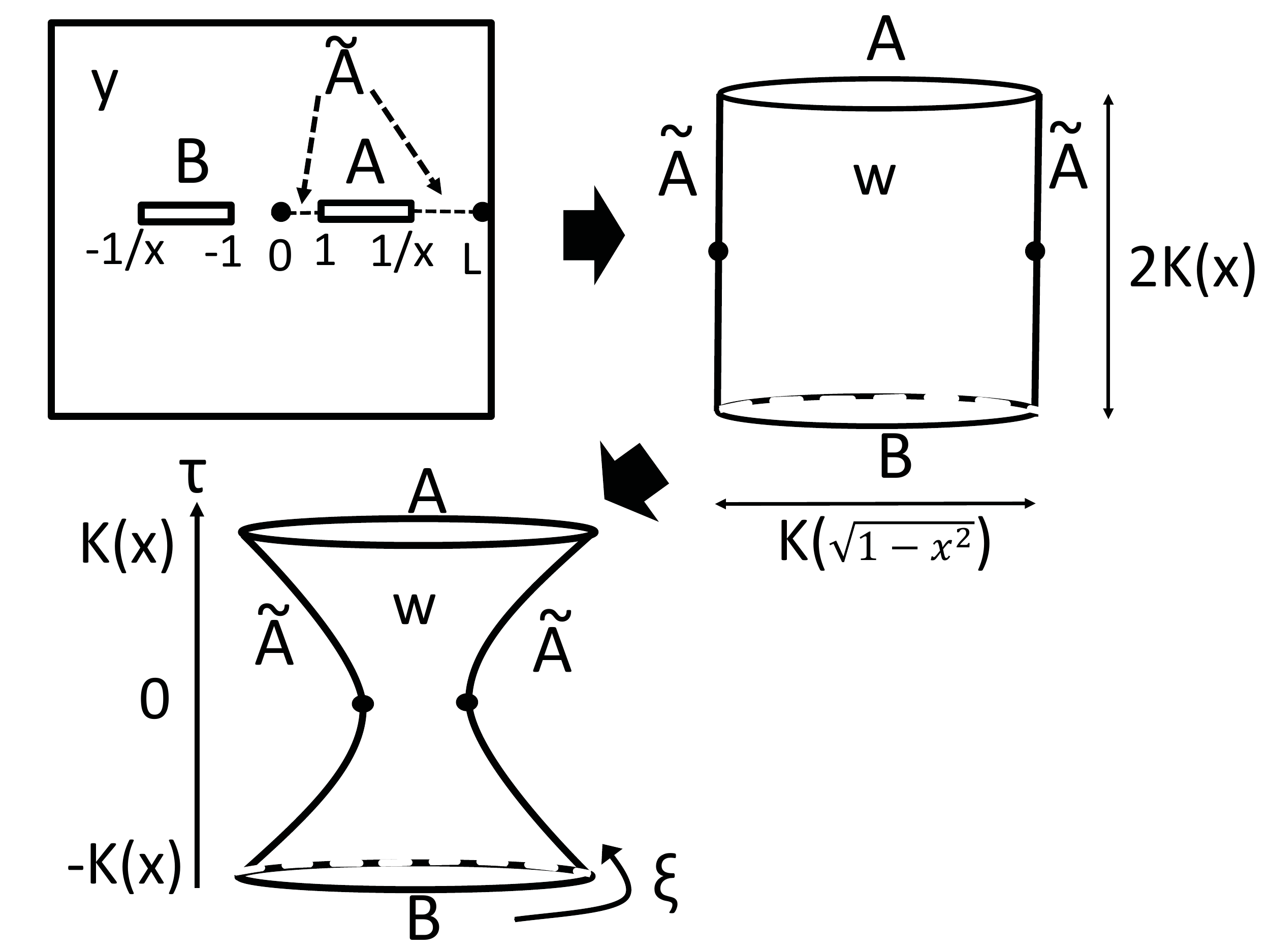}
 \caption{Conformal map from complex plane with two cuts to a cylinder and its path-integral optimization.}
\label{fig:Conformal}
\end{figure}

The reduced density matrix $\rho_{AB}$ is described by an Euclidean path-integral on the complex plane ($y$-plane) with two slits along $A$ and $B$ as in the upper left picture of Fig.\ref{fig:Conformal}.
Since $\rho_{AB}$ depends only on the cross ratio $\eta$ owing to the conformal invariance,
without losing generality, we can fully parameterize the subsystem $A$ and $B$ as:
\ba
A=[1,1/x],\ \ \ B=[-1/x,-1],\ \ \ (0<x<1).
\ea
The cross ratio is given by $\eta=(1-x)^2/4x$. We can choose the subsystem $\ti{A}$  and
$\ti{B}$ which purify $\rho_{AB}$ such that
 $A\ti{A}=[0,L]$ and take the limit $L\to\infty$, owing to the
 reflection symmetry. We confirmed numerically that this reflection symmetry is not spontaneously broken.

The $y$-plane with two slits is mapped into a cylinder, which is described by the coordinate
$w=\xi+i\tau$, as:
\be
w=i\int^y_0 d\ti{y}\frac{1}{\s{(1-\ti{y}^2)(1-x^2\ti{y}^2)}}, \label{ww}
\ee
as in the upper right of Fig.\ref{fig:Conformal}.

Finally, we perform the path-integral optimization on a cylinder as in the lower picture of Fig.\ref{fig:Conformal}. We obtain the optimized metric by solving the Liouville equation:
\be
ds^2=e^{2\phi}dw d\bar{w}=e^{2\ti{\phi}}dyd\bar{y},
\label{wtr}
\ee
where
\ba
&& e^{2\phi}=\frac{\pi^2}{4K(x)^2}\cdot \frac{\ep^2}{\cos^2\left(\frac{\pi\tau}{2K(x)}\right)},\no
&& e^{2\ti{\phi}}=e^{2\phi}\cdot \frac{1}{|(1-y^2)(1-x^2y^2)|^2}. \label{opdbw}
\ea
We introduced the elliptic function $K(x)$:
\be
K(x)=\int^1_0 d\ti{y}\frac{1}{\s{(1-\ti{y}^2)(1-x^2\ti{y}^2)}}.
\ee

The coordinates $\tau$ takes values in the range:
\be
-K(x)+\delta_2\leq \tau \leq K(x)-\delta_2,  \label{regua}
\ee
and $\xi$ is periodically identified: $\xi\sim \xi+2K(\s{1-x^2})$. Here $\delta_2$ is an infinitesimally small regularization parameter fixed by the condition (\ref{condw}).
Again, the detail of $\delta_2$ is not important in our calculations below and refer to appendix A
for details.
The boundary points of $A\ti{A}$: $y=0$ and $y=L(\to\infty)$ are mapped into the antipodal two points in the middle: $(\xi,\tau)=(0,0)$ and $(\xi,\tau)=(K(\s{1-x^2}),0)$.

Now the entanglement entropy $S^{min}_{A\ti{A}}$ after the optimization
can be computed as follows:
\ba
S^{min}_{A\ti{A}}&=&\frac{c}{3}\log \frac{L}{\ep}+\frac{c}{6}\ti{\phi}|_{y=L}+\frac{c}{6}\ti{\phi}|_{y=0}\no
&=&\frac{c}{3}\log \frac{L}{\ep}
+\frac{c}{6}\log\left[\frac{\pi\ep}{2L^2xK(x)}\right]
+\frac{c}{6}\log\left[\frac{\pi\ep}{2K(x)}\right]\no
&=&-\frac{c}{6}\log x -\frac{c}{3}\log\left[\frac{2K(x)}{\pi}\right].
\label{optdb}
\ea

Since $K(x)\simeq \frac{\pi}{2}+\frac{\pi}{8}x^2+\ddd$ for $x\ll 1$, we have
\be
S^{min}_{A\ti{A}}=-\frac{c}{6}\log x +O(x^2).
\ee

We can compare this with the holographic EoP \cite{UT,Nguyen:2017yqw}:
\ba
E_W(\rho_{AB})=-\frac{c}{6}\log x, \ \ \   (0<x<3-2\s{2}).
\ea
For $3-2\s{2}<x<1$ we have $E_W(\rho_{AB})=0$ as the entanglement wedge gets disconnected.

\begin{figure}[ttt]
  \centering
 \includegraphics[width=6cm]{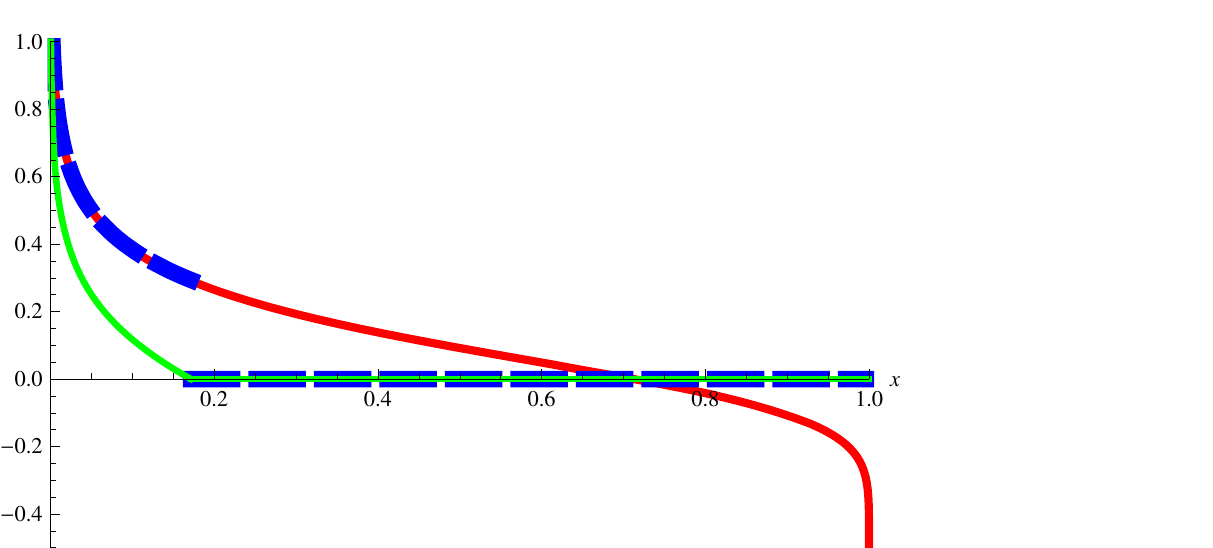}
 \caption{Plots of HEoP (red), $I(A:B)/2$ (green) and $S^{min}_{A\ti{A}}$ (blue dashed) as a function of
 $x$. The first two get vanishing for $x>3-2\s{2}$. Note that the plot of $S^{min}_{A\ti{A}}$ is valid only 
when $x\ll 1$, where it coincides with HEoP.}
\label{fig:plot1}
\end{figure}

It is also useful to note that the holographic
mutual information (HMI) $I(A:B)=S_{A}+S_B-S_{AB}$ behaves
\be
\frac{1}{2}I(A:B)=\frac{c}{6}\log z =\frac{c}{6}\log \frac{(1-x)^2}{4x}\ \ \ \   (0<x<3-2\s{2}),
\ee
while $I(A:B)=0$ for $3-2\s{2}<x<1$.
We plotted these quantities in Fig.\ref{fig:plot1}.

We notice that $S^{min}_{A\ti{A}}$ almost coincides with $E_W(\rho_{AB})$ for $0<x<3-2\s{2}$ in the plot.
However, one may worry that they deviate from each other at the order $O(x^2)$. Also,
$S^{min}_{A\ti{A}}$ continuously decreases and becomes incorrectly negative as $x$ gets larger.
We would like to argue that our Weyl invariance (\ref{wtr}) breaks down except for $x\ll 1$.
This is because the length of minimal cross section of the optimized cylinder, given by $L_{cyl}=\frac{\pi\ep K(\s{1-x^2})}{K(x)}$, is much greater than the cut off scale $\ep$ only if $x\ll 1$. If the size of manifold
on which we path-integrate gets the same order of the lattice spacing $\ep$,  we cannot trust field theoretic properties of path-integrals, such as the Weyl invariance. Therefore, the result (\ref{optdb})
is trustable only for $x\ll 1$ and in this range the equality $S^{min}_{A\ti{A}}=E_W(\rho_{AB})$ is confirmed.
Indeed, in the single interval case, where we found the perfect matching (\ref{eopmin}), the Weyl transformation is trustable for any values of $a,b$ and $p$ because we always have $L_{cyl}=\infty$.

It is also useful to note that at the holographic phase transition point $x=3-2\s{2}$ (or equally $\eta=1$),
the minimum cross section of the cylinder becomes $L_{cyl}=2\pi \ep$. This indeed agrees with the point where
a counterpart of confinement/deconfinement transition is expected in the path-integral optimization
for holographic CFTs as argued in \cite{Caputa:2017urj}, though we cannot fully trust this argument because
$L_{cyl}$ gets as small as the lattice spacing $\ep$.

\vspace{5mm}

{\bf 5. Conclusions and Discussions}

In this letter, for a given mixed state $\rho_{AB}$, we introduced a minimized entanglement entropy $S^{min}_{A\ti{A}}$ as follows. First we define a special purified state $|\Psi_{min}\lb_{ABC}$ for $\rho_{AB}$ by minimizing the path-integral complexity $C_L$ (\ref{pco}). Then $S^{min}_{A\ti{A}}$ is given
by taking the minimum of $S_{A\ti{A}}$ for the purified state $|\Psi_{min}\lb_{ABC}$ among all possible
decompositions of the ancilla space for purification $C=\ti{A}\cup \ti{B}$.

We analytically computed this quantity $S^{min}_{A\ti{A}}$ in several setups in two dimensional CFTs. Remarkably, we found that it matches with the area of the entanglement wedge cross section $E_W(\rho_{AB})$ in AdS, called the holographic entanglement of purification.

It is not obvious at present if our special purification is identical to the one which minimizes $S_{A\ti{A}}$ as in the original definition (\ref{EoPdef}). However, we expect they are at least very close to each other because the former minimizes the path-integral complexity and thus efficiently compresses the sizes of $\ti{A}$ and $\ti{B}$. In this sense, our result can be regarded as the first quantitative evidence for the holographic
EoP conjecture \cite{UT,Nguyen:2017yqw}.

In the original argument \cite{UT}, the equivalence between $E_W(\rho_{AB})$ and $E_P(\rho_{AB})$ was explained by assuming the surface/state duality \cite{MT}, based on the conjectured relation between tensor networks \cite{MERA,TNR} and AdS/CFT \cite{Swingle}. In this argument, the minimization is taken only for quantum states with classical gravity duals. We would like to leave for future works the precise meaning of the minimum in the EoP (\ref{EoPdef}) which correctly matches with $E_W(\rho_{AB})$.  Higher dimensional 
generalizations are also an intriguing future problem.\\

{\bf Acknowledgements} We thank Juan Maldacena, Brian Swingle, Kotaro Tamaoka and Guifre Vidal for useful conversations.  MM and KU are supported by JSPS fellowships. PC and TT are supported by the Simons Foundation through the ``It from Qubit'' collaboration. PC is supported by JSPS Grant-in-Aid for Research Activity start-up 17H06787. MM is supported by JSPS Grant-in-Aid for JSPS Fellows No.16J08909. TT is supported by JSPS Grant-in-Aid for Scientific Research (A) No.16H02182 and by JSPS Grant-in-Aid for Challenging Research (Exploratory) 18K18766. TT is also supported by World Premier International Research Center Initiative (WPI Initiative) from the Japan Ministry of Education, Culture, Sports, Science and Technology (MEXT). KU is supported by JSPS Grant-in-Aid for JSPS Fellows No.18J2288
8.

\vspace{5mm}

\section{Appendix A: Details of Regularization}

In the path-integral optimization procedures, we encountered metrics $ds^2=e^{2\ti{\phi}}dyd\bar{y}$ which get divergent at the endpoints of $AB$ such as (\ref{metsop}) for the single interval and (\ref{opdbw}) for the double intervals. To maintain the boundary condition
(\ref{condw}) on the slit $AB$, we actually need to shift the location of $AB$ which is infinitesimal in the continuum limit $\ep\to 0$. This is equivalent to the cut off of the range of $\tau$ coordinate, denoted by
 $\delta_1$ and $\delta_2$ in the main context of this letter for the single and double interval, respectively (refer to e.g.(\ref{regua}) for the latter).

\begin{figure}[ttt]
  \centering
 \includegraphics[width=8cm]{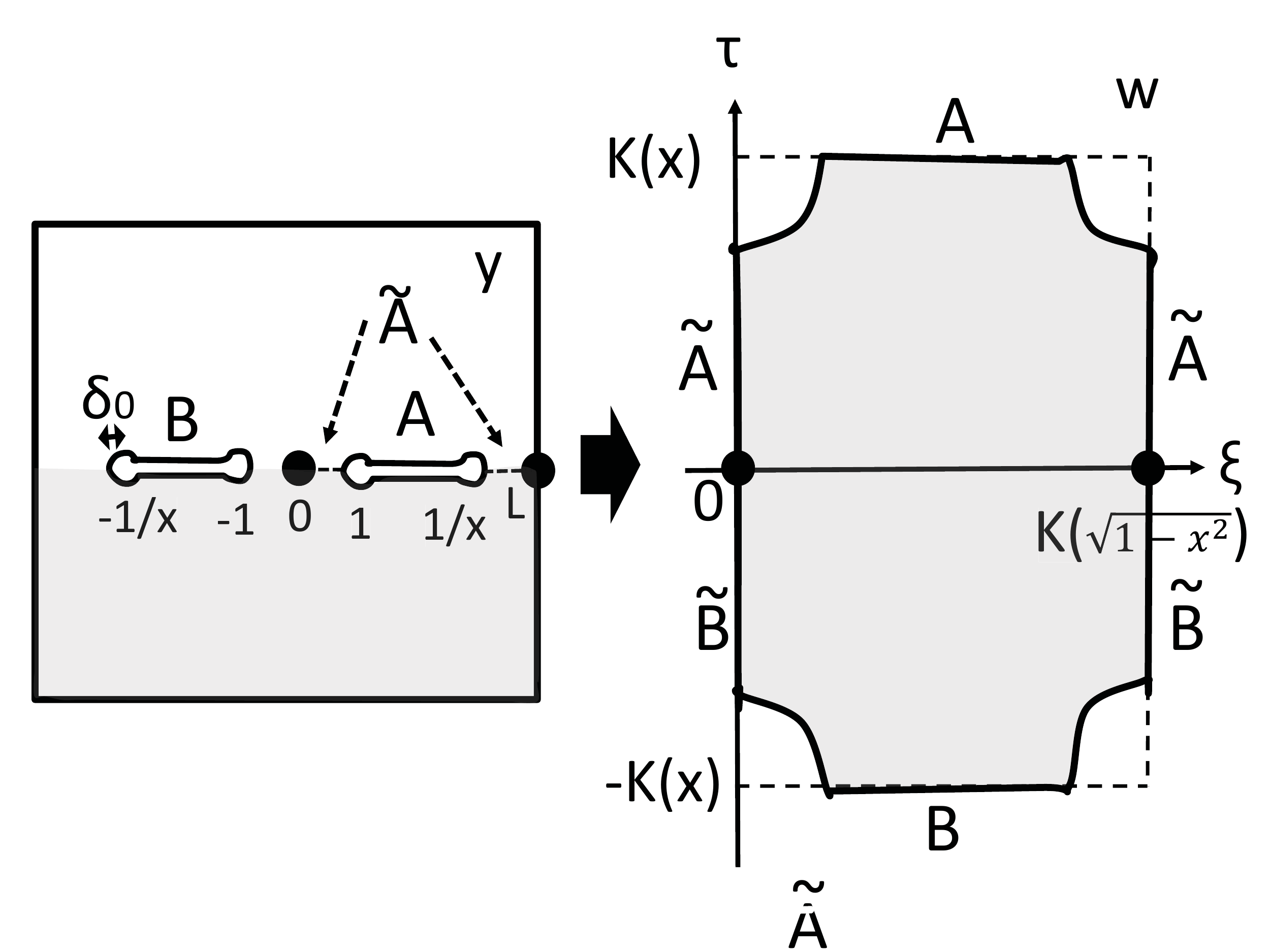}
 \caption{A sketch of regularized conformal transformations, when $AB$ are double intervals in the CFT vacuum. The grey colored region in the left is mapped to the
 same colored region in the right. The total plane with the two cuts on $A$ and $B$ eliminated, is mapped into two copies of the right picture, i.e. a cylinder. }
\label{fig:Conformalreg}
\end{figure}

Let us study the detail of such a cut off by focusing on the double interval case, as the single interval case can be treated almost in the same way. The boundary condition (\ref{condw}) fixes the position dependent regularization $\delta_2$ in (\ref{regua}) as
\be
\delta_2=\frac{\ep}{\s{|(1-y^2)(1-x^2y^2)|}}. \label{delk}
\ee
One may worry that the relation (\ref{delk}) gets singular at the boundaries of $A$ i.e.
$y=1$ and $y=1/x$. However, this is not a problem as we can regularize the divergences at the two end points by eliminating infinitesimally small neighborhoods
around them: $|y-1|\leq \delta_0$ and  $|y-1/x|\leq \delta_0$ such that $\delta_2$ is
always small. Then we apply the same conformal map (\ref{ww}) to this regularized region.
This is sketched in Fig.\ref{fig:Conformalreg}. The regularization corresponds to eliminating the four edges of the cylinder in the $w$ coordinate. Since we can take $\delta_0$ infinitesimally small, these removed edges also get infinitesimally small.
 Because the twist operators are inserted at the end points of $A\ti{A}$, i.e.
 $w=0$ and $w=K(\s{1-x^2})$, this regularization procedure does not affect our calculation of $S^{min}_{A\ti{A}}$.

\end{document}